\documentclass{rnaastex}

\usepackage{graphics,graphicx,url,verbatim,xspace}
\usepackage{glossaries}
\usepackage{comment}
\usepackage{makecell}
\usepackage{multirow}

\definecolor{black}{rgb}{0,0,0}
\definecolor{red}{rgb}{1.0,0,0}
\newcommand{\UCB}{Department of Astronomy,  University of California, Berkeley, Berkeley, CA 94720}

\newcommand{\SWIN}{Centre for Astrophysics \& Supercomputing, Swinburne University of Technology, Hawthorn, VIC 3122, Australia}

\newcommand{\SETI}{SETI Institute, Mountain View, California}
\newcommand{\KZA}{University of Malta, Institute of Space Sciences and Astronomy}
\newcommand{\BPF}{The Breakthrough Initiatives, NASA Research Park, Bld.~18, Moffett Field, CA, 94035, USA}
\newcommand{\PENN}{Department of Astronomy and Astrophysics, Pennsylvania State University, University Park PA 16802}

\begin{document}

\title{Breakthrough Listen Search for Technosignatures Towards the Kepler-160 System}


\correspondingauthor{Karen Perez}
\email{karen.i.perez@columbia.edu}


\author{Karen Perez}
\affiliation{Department of Astronomy, Columbia University, 550 West 120th Street, New York, NY 10027, USA}

\author{Bryan Brzycki}
\affiliation{\UCB}

\author[0000-0002-8604-106X]{Vishal Gajjar}
\affiliation{\UCB}

\author[0000-0002-0531-1073]{Howard Isaacson}
\affiliation{\UCB}
\affiliation{Centre for Astrophysics, University of Southern Queensland, Toowoomba, QLD, Australia}

\author[0000-0003-2828-7720]{Andrew Siemion}
\affiliation{\UCB}
\affiliation{\SETI}
\affiliation{\KZA}

\author[0000-0003-4823-129X]{Steve Croft}
\affiliation{\UCB}
\affiliation{\SETI}

\author[0000-0003-3197-2294]{David DeBoer}
\affiliation{\UCB}

\author{Matt Lebofsky}
\affiliation{\UCB}

\author{David H.\ E.\ MacMahon}
\affiliation{\UCB}

\author[0000-0003-2783-1608]{Danny C.\ Price}
\affiliation{\UCB}
\affiliation{\SWIN}

\author{Sofia Sheikh}
\affiliation{\PENN}

\author{Jamie Drew}
\affiliation{\BPF}

\author{S. Pete Worden}
\affiliation{\BPF}


\begin{abstract}
We have conducted a search for artificial radio emission associated with the Kepler-160 system following the report of the discovery of the Earth-like planet candidate KOI-456.04 on 2020 June 4 \citep{Heller:2019}.  Our search targeted both narrowband (2.97\,Hz) drifting ($\pm 4$\,Hz~s$^{-1})$ and wideband pulsed (5\,ms at all bandwidths) artificially-dispersed technosignatures using the turboSETI \citep{Enriquez:2017} and SPANDAK \citep{SPANDAKinprep} pipelines, respectively, from 1$-$8 GHz. No candidates were identified above an upper limit Equivalent Isotropic Radiated Power (EIRP) of $5.9 \times 10^{14}$ W for narrowband emission and $7.3 \times 10^{12}$ W for wideband emission.  Here we briefly describe our observations and data reduction procedure.

\end{abstract}


\section{Introduction}

\label{sec:intro}

\cite{Heller:2019} recently identified a previously unknown nontransiting planet in the Kepler-160 (KOI-456) system (Kepler-160d), as well as a super-Earth-sized transiting planet candidate, KOI-456.04 based on the $25^{th}$ \emph{Kepler} Data Release for \emph{Kepler} Quarters 1 -- 17.  These new discoveries add to the two already known transiting planets orbiting the Sun-like (1.12 R\textsubscript{\(\odot\)} and $T_{eff}$ = 5471 K) star Kepler-160. In their search for new transiting planets to explain the periodic transit-timing variations (TTVs) exhibited by Kepler-160c, the planet candidate with 1.91 Earth radii, an orbital period of 378 days, and an estimated surface temperature of 245\,K, was found. The TTVs of Kepler-160c can be explained by the new nontransiting planet, Kepler-160d, but these TTVs are unrelated to KOI-456.04. This Earth-like planet candidate, given its ideal location in the habitable zone of its host star and edge-on orientation projected toward Earth, represents an ideal target for technosignature searches.

Here, we describe observations and analysis of the Kepler-160 system using the Green Bank Telescope (GBT) as part of the ongoing Breakthrough Listen search for technosignatures \citep{Worden:2017, Isaacson:2017,gaj20}. We searched the radio frequency bands 1.10 -- 1.90\,GHz (L-band), 1.80 -- 2.80\,GHz (S-band), and 3.95 -- 8\,GHz (C-band) for narrowband (2.97\,Hz) Doppler-accelerated ($\pm 4$\,Hz~s$^{-1})$ and wideband (5\,ms at all bandwidths) artificially-dispersed technosignatures using the turboSETI \citep{Enriquez:2017} and SPANDAK \citep{SPANDAKinprep} pipelines, respectively. 


\section{Observations}
\label{sec:obs}

We observed Kepler-160 with the GBT for three ON-OFF 5-minute pointings at each frequency band, beginning on UT 2020 June 14 11:13:36 with L-band, using the J2000 ephemeris (19$^h$\,11$^m$\,05.52$^s$, 42\degr\,52\arcmin\,19\farcs12). The table below shows the start and end times for observations at each frequency band on 2020 June 14. Observations were recorded using the Breakthrough Listen backend \citep{MacMahon:2018}. A total of 18 observations were made: 3 on-target and 3-off target observations per frequency band.

Raw voltage data were reduced to multiple data products using the standard Listen data reduction pipeline \citep{Lebofsky:2019}. All observational data used in this note are available online, including all three standard BL data products (bandwidth / integration time of $\sim 2.79$\,Hz / $\sim 18.25$\,s, $\sim 366.21$\,kHz / $\sim 349.53$\,$\mu$s and $\sim 2.86$\,kHz / $\sim 1.07$\,s) for each pointing\footnote{\url{http://blpd1.ssl.berkeley.edu/kepler160/}}. The observations used in this analysis are labeled KEPLER-160. There are 3 sets, each corresponding to a different frequency band, and each set is separated by other observations. The first L-band observation, numbered 0010, is an on-target pointing, after which on- and off-target pointings alternate.  The first S-band observation is numbered 0025, and the first C-band observation is numbered 0037. The on and off-target alternations apply here as well. The high spectral resolution, high time resolution, and medium resolution data products have suffixes of 0000, 0001, and 0002, respectively. For SETI searches, we use the high spectral resolution product (0000) while for the transient searches we used high time resolution (0001) products.

Table~\ref{tab:table1} shows our observational parameters for each frequency band. The mean system temperature is also shown, determined from observations of 3C\,147 for L- and S-bands, and 3C\,196 for C-band.
    
\begin{table}[h]
\caption{Observational Parameters and Calculated Values for each frequency band}
\begin{center}
    \begin{tabular}{c c c c c c c c }
    \hline
    \hline
    \multicolumn{8}{c}{GBT bands} \\
    \hline
    Band & {Frequency} & {Start Time} & {End Time} & Measured & SEFD & {10$\sigma$ Detectable Flux} & {10$\sigma$ EIRP limit} \\
    & Range (GHz) & (UTC) & (UTC) & $T_\textrm{sys}$ (K) & (Jy) & Density Limit (Jy) & {($10^{12}\,W$)}\\
    \hline
    L-band & 1.10 -- 1.90 & 11:13:36 & 11:40:07 & 24.7 & 12.4 & 8.5 & 890 \\
    S-band & 1.80 -- 2.80 & 12:21:09 & 12:47:37 & 19.7 & 9.9 &  6.7& 710 \\
    C-band & 3.95 -- 8.00 & 13:45:45 & 14:12:12 & 16.6 & 8.3 & 5.7 & 600 \\

    \hline
    \end{tabular}
    \label{tab:table1}
\end{center}
\end{table}
 
\section{Results and Discussion}

To search for narrowband signals, we used the turboSETI code \citep{Enriquez:2017} over a Doppler drift rate range of $\pm 4$\,Hz~s$^{-1}$ and a signal-to-noise (S/N) threshold of 10 with the $\sim$2.79\,Hz / $\sim$18.25\,s data product, broadly following the steps described by \citet{Price_2020} and \citet{Sheikh_2020}. Any continuous narrowband signals associated with pointings toward Kepler-160 would appear in all 3 of the ON observations and in none of the OFF observations. More candidates are detected at L and S bands, which is consistent with high RFI levels at lower frequencies; however, all potential candidates appear in at least one OFF observation, which rules out any spatially isolated signals in the direction of Kepler-160. 

Using Kepler-160's distance of $3141_{-54}^{+56}$ light years \citep{Gaia2016, Gaia2018, Andrae_2018} and Equations 3 -- 4 of \cite{Enriquez:2017} and our calibrated system equivalent flux density (SEFD), we calculated the minimum detectable flux density for a 300\,s observation and a $10\sigma$ threshold for each frequency band. Using these values and \citeauthor{Enriquez:2017} Equations 5 -- 7, we derived the upper limit for the EIRP of a hypothetical transmitter. These values, calculated for each frequency band, are also shown in Table~\ref{tab:table1}.

We also conducted a comprehensive search for transient signals with artificial dispersion \citep{lgw+20} using the SPANDAK\footnote{\url{https://github.com/gajjarv/PulsarSearch}} GPU-accelerated transient detection pipeline operating on the $\sim$366.21\,kHz / $\sim$349.53\,$\mu$s data product.  We did not find any transient candidates from the observations listed in Table~\ref{tab:table1}. We estimate fluence limits of 0.2, 0.18, and 0.07\,Jy\,ms for transient bursts with S/N~$ = 10$ and width 5\,ms at L, S, and C bands, respectively.  The corresponding EIRP limits are $2.1 \times 10^{13}$\,W, $1.9 \times 10^{13}$\,W, and $7.3 \times 10^{12}$\,W, respectively. 

Kepler-160 was also observed at optical wavelengths on the Keck~I telescope with HIRES on UT 2014 August 19 \citep{Petigura_2017}. \citet{tellis2015search} searched this and other Kepler host stars for optical laser lines; however, they found no artificial optical laser emission. The laser emission detectability limit for a typical star in their analysis is $\sim 10^{-2}$ photons m$^{-2}$~s$^{-1}$. Several candidate laser lines that rise above the stellar continuum can be attributed to cosmic rays upon inspection of the raw image.

Future observations of KOI-456.04 with future missions, like PLATO \citep{2014ExA....38..249R}, might recover its transit, confirming its candidacy as a planet and aiding in any further radio observations. Additionally, we expect to carry similar searches towards other exoplanet systems and candidates as more targets of interest are discovered using ongoing missions such as TESS and K2. 

\acknowledgments

Breakthrough Listen is managed by the Breakthrough Initiatives, sponsored by the \href{http://breakthroughinitiatives.org}{Breakthrough Prize Foundation}.  The Green Bank Observatory is a facility of the National Science Foundation, operated under cooperative agreement by Associated Universities, Inc.

\bibliographystyle{aasjournal}
\vspace{1 cm}
\bibliography{references}

\end{document}